\documentclass[pdflatex,sn-mathphys-num]{sn-jnl}

\usepackage{graphicx}%
\usepackage{multirow}%
\usepackage{amsmath,amssymb,amsfonts}%
\usepackage{amsthm}%
\usepackage{mathrsfs}%
\usepackage[title]{appendix}%
\usepackage{xcolor}%
\usepackage{textcomp}%
\usepackage{manyfoot}%
\usepackage{booktabs}%
\usepackage{algorithm}%
\usepackage{algorithmicx}%
\usepackage{algpseudocode}%
\usepackage{listings}%
\usepackage{indentfirst}
\usepackage{verbatim}

\usepackage{booktabs}
\usepackage{threeparttable}

\begin{document}

\title[Article Title]{Computing band gaps of periodic materials via sample-based quantum diagonalization}


\author[1,7]{\fnm{Alan} \sur{Duriez}}
\equalcont{These authors contributed equally to this work.}

\author[2,9]{\fnm{Pamela} \sur{C. Carvalho}}
\equalcont{These authors contributed equally to this work.}

\author[1,8]{\fnm{Marco Antonio} \sur{Barroca}}
\equalcont{These authors contributed equally to this work.}

\author[3,10]{\fnm{Federico} \sur{Zipoli}}

\author[4]{\fnm{Ben} \sur{Jaderberg}}

\author[1]{\fnm{Rodrigo} \sur{Neumann Barros Ferreira}}

\author[5]{\fnm{Kunal} \sur{Sharma}}

\author[5]{\fnm{Antonio} \sur{Mezzacapo}}

\author[6]{\fnm{Benjamin} \sur{Wunsch}}

\author*[1]{\fnm{Mathias} \sur{Steiner}}\email{mathiasbsteiner@gmail.com}

\affil[1]{\orgname{IBM Research}, \orgaddress{\city{Rio de Janeiro}, \postcode{20031-170}, \state{RJ}, \country{Brazil}}}

\affil[2]{\orgname{IBM Research}, \orgaddress{\city{São Paulo}, \postcode{04007-900}, \state{SP}, \country{Brazil}}}

\affil[3]{\orgname{IBM Research Europe}, \orgaddress{\city{Zurich}, \postcode{8803}, \state{Rüschlikon}, \country{Switzerland}}}

\affil[4]{\orgname{IBM Research Europe}, \orgaddress{\city{Hursley, Winchester}, \postcode{SO21 2JN}, \country{United Kingdom}}}

\affil[5]{\orgname{IBM Quantum}, \orgname{IBM T.J. Watson Research Center, Yorktown Heights}, \orgaddress{\city{New York}, \postcode{10598}, \state{NY}, \country{USA}}}

\affil[6]{\orgname{IBM Research}, \orgname{IBM T.J. Watson Research Center, Yorktown Heights}, \orgaddress{\city{New York}, \postcode{10598}, \state{NY}, \country{USA}}}

\affil[7]{\orgdiv{Instituto de Física}, \orgname{Universidade Federal Fluminense}, \orgaddress{\city{Niterói}, \postcode{24210-346},\state{RJ}, \country{Brazil}}}

\affil[8]{ \orgname{Centro Brasileiro de Pesquisas Físicas}, \orgaddress{\city{Rio de Janeiro}, \postcode{22290-180}, \state{RJ}, \country{Brazil}}}

\affil[9]{\orgdiv{Instituto de Física da Universidade de São Paulo},\orgname{Universidade de São Paulo}, \orgaddress{\city{São Paulo}, \postcode{05508-090}, \state{SP}, \country{Brazil}}}

\affil[10]{\orgname{National Center for Competence in Research-Catalysis
(NCCR-Catalysis)}, \orgaddress{\city{Zurich}, \country{Switzerland}}}


\abstract{A key objective of computational solid state physics is to predict electronic properties of periodic materials. However, electronic structure simulations based on density functional theory fail to predict experimental results if correlations are not properly accounted for. Here, we report a sample-based quantum diagonalization workflow for simulating electronic states of periodic materials, and for predicting their band gaps. To that end, we devise a general lattice Hamiltonian representation in which material-specific, electronic interaction parameters are obtained self-consistently. Two exemplar, wide-gap materials - hafnium dioxide and zirconium dioxide - are expressed as quantum circuits that leverage the lattice representation with a materials-specific parametrization. We sample the quantum circuits on a state-of-the-art, superconducting quantum processor and diagonalize the lattice Hamiltonian in the reduced configuration subspaces with standard techniques. Our method outperforms select quantum-chemical benchmarks as well as approaches based on density functional theory, the standard reference in materials simulation of solids. Importantly, the quantum-computed band gap predictions for the two dielectrics agree with independent lab experiments. In essence, quantum-classical hybrid simulation workflows on pre-fault tolerant quantum computers produce useful, experimentally verifiable property predictions in applied  materials science.}




\maketitle


Considerable effort in the fields of computational chemistry and materials science is dedicated to property predictions \cite{Alexeev_2024}. A reference approach to electronic structure predictions is based on the Kohn-Sham (KS) construction \cite{Kohn1965} in density-functional theory (DFT) \cite{Hohernberg1964}.  The method has been successfully applied in the investigation of metals, semiconductors and insulators \cite{Burke2013}. However, the mean-field approximations and assumption of non-interacting electrons are not justified in transition-metal oxides \cite{Mandal2019}, where atomic-scale interactions of ions with localized, open \textit{d} and \textit{f} shells are preserved in extended, crystalline bulks \cite{Pavarini2021, Solovyev2008}. The occurrence in cuprates of  high-temperature superconductivity \cite{Bussmann2020} and Mott insulation \cite{Sachdev2003}, for example, is due to the strong electronic correlations that are neglected in standard, DFT-based simulations.  

In periodic materials with strong electronic interactions, the prediction of the energy barrier between the highest occupied and lowest unoccupied electronic states, i.e., the band gap, is particularly challenging \cite{Pavarini2021}. The predicted E$_{\textnormal{gap}}$-values typically underestimate the experimental values obtained for wide-gap semiconductors \cite{Schluter1990,Pavarini2021}. The accuracy of DFT-based predictions can be improved by including energy corrections for \emph{d}- and \emph{f}-electrons. In the Extended Hubbard (EH) model  \cite{Zhang1989}, strong electronic correlations are captured by the intra-site and inter-site parameters U and V, respectively. Typically, U corrects for contributions of localized atomic orbitals in transition metals and V represents the interaction between metal ions and their nearest neighbors. Inclusion of Hubbard parameters in EH representation, which can be derived from first principles \cite{Campo2010, Timrov2022}, improves band gap prediction accuracy with respect to tight-binding (TB) approaches \cite{himmetoglu2014hubbard}. Nevertheless, the predictions might still fall short of experimental results \cite{Naveas2023}.

\begin{figure}[!htb] 
\includegraphics[width=\textwidth]{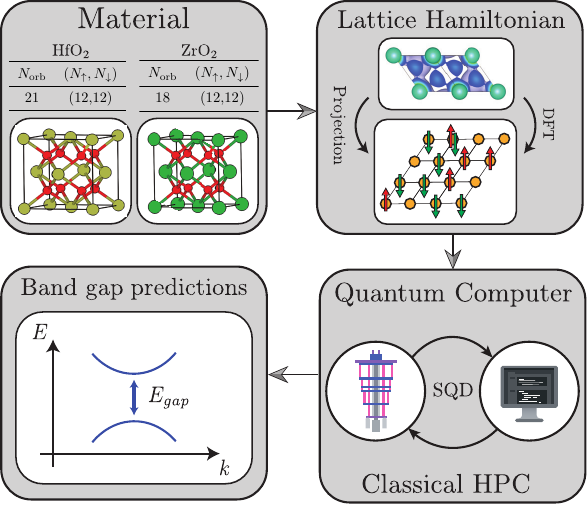} 
    \caption{Hybrid, quantum-classical computational workflow for predicting band gaps of periodic materials. Based on a lattice Hamiltonian representation with materials-specific parametrization, sample-based quantum algorithms compute band gap energies for comparison with experimental results.}
    \label{fig:workflow}
\end{figure}

In the EH framework, the electronic structure is expressed by a fermionic lattice Hamiltonian. Successful model representations have been reported for a variety of applied materials, among them  organic conductors  \cite{kivelson1987, castet1996}, twisted bilayer graphene \cite{liao2019} and high-temperature superconductors \cite{luo2025}. Numerical simulations of lattice models carry, however, substantial computational cost and are, therefore, limited to very small systems \cite{liang1994}. Both Quantum Monte Carlo (QMC) \cite{ceperlay1980} and Density Matrix Renormalization Group (DMRG) \cite{white1992, liang1994, sobcak2009}, have been employed to approximate the ground state energies of lattice systems.  Nevertheless, while QMC has limited applicability to fermionic systems \cite{troyer2005,pan2024}  DMRG is difficult to scale to three dimensions \cite{Mezzacapo2009}.

Quantum computing opens a route to simulating lattice representations of periodic materials at scale. Hamiltonian problem formulations are straightforward to implement with today's quantum algorithms \cite{Alexeev_2024} and are among the candidates for demonstrating a provable quantum advantage \cite{Wu2024}. While variational quantum algorithms were tested for predicting the band structure of bulk silicon \cite{Ohgoe2024}, the size limitations and noise levels of current quantum processors have so far limited their applicability \cite{Zhao2025}. 

Recently, the sample-based quantum diagonalization (SQD) method \cite{chemistry-paper} was applied to the simulation of the electronic properties of molecular structures. Within a quantum-centric workflow, SQD integrates the capabilities of quantum computers with classical, high-performance computing (HPC) environments for predicting electronic energies of molecular systems represented by as much as 72 spin orbitals for comparison with classical, quantum-chemical reference methods \cite{chemistry-paper}. The SQD algorithm samples a molecule's electronic configuration space that is represented by a quantum circuit with material-specific parametrization. This can also be understood as quantum-selected configuration interaction \cite{Kanno2023, Ohgoe2024}.  

The method of diagonalizing a quantum-selected, electronic configuration subspace could, in principle, be applied to periodic material represented by an EH Hamiltonian. This would require to map a material's EH lattice Hamiltonian onto a generalized electronic Hamiltonian suitable for quantum sampling. The quantum circuit could be implemented by means of a local unitary cluster Jastrow ansatz \cite{lucj-ansatz}, with a parameter initialization taken from a coupled-cluster singles and doubles (CCSD) calculation \cite{chemistry-paper}. If successful, the approach would enable band gap predictions with higher accuracy than DFT-based simulations.

\begin{figure}[!htb] 
    \centering
    \hspace{-0.6cm}
    \includegraphics[width=1.\textwidth]{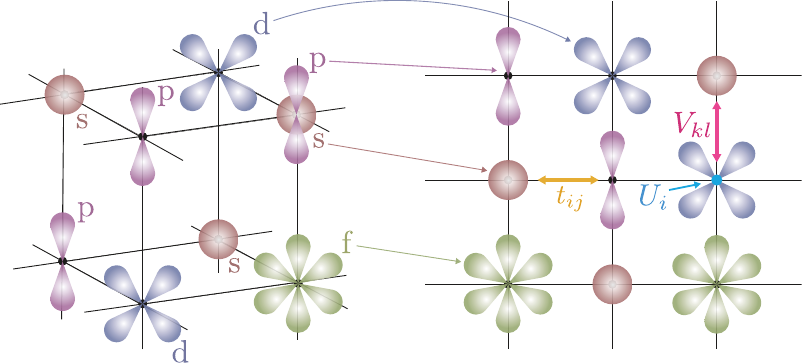} 
    \caption{Lattice Hamiltonian representation of periodic materials. A tight-binding Hamiltonian with hopping parameters $t_{ij}$ is complemented with intra-site $U_{i}$ and inter-site $V_{kl}$ interaction parameters, see \autoref{eq:hubbard-hamiltonian}, that are calculated self-consistently for a specific material within the DFT+U+V framework. The electronic density is projected onto a localized, atomic orbital basis set. Each atomic orbital of every atom in the crystal is mapped onto a single site in the projected lattice, as illustrated here for a representative solid with a cubic lattice.}
    \label{fig:projection}
\end{figure}

In the following, we introduce a general EH representation of the electronic structure of periodic materials for predicting their band gaps using a quantum computer. As visualized in \autoref{fig:workflow}, the lattice representation is obtained by projecting DFT-based, electronic densities onto a localized atomic orbital basis set, while U- and V- corrections are computed self-consistently. We then apply the SQD method for computing the ground state energy of the customized EH Hamiltonian, using CCSD results for initial parametrization. To validate the method for application in material science, we benchmark the SQD-based band gap predictions for two representative dielectrics, hafnium dioxide (HfO\textsubscript{2}) and zirconium dioxide (ZrO\textsubscript{2}), against results obtained with DFT simulations and standard, quantum-chemical reference methods. While outperforming DFT+U+V as well as CCSD by wide margins, we find that the SQD method predicts band gaps well within the range of reported lab results.  


The construction of a general EH lattice Hamiltonian for periodic materials requires a projection of the DFT electronic structure onto a localized basis set for obtaining a tight-binding Hamiltonian with identical band structure, see Methods section and Supporting Information for details.  We select the valence atomic orbitals in the DFT pseudopotential as reference basis set, see Methods section. As visualized in \autoref{fig:projection}, each atomic orbital of every atom within the crystal is mapped onto a single site in the projected lattice. Note, that orbitals of same atom are mapped onto different lattice sites.

To account for electronic correlations within the lattice representation, we add intra-site and inter-site Hubbard interaction parameters, respectively, that are computed self-consistently via Density Functional Perturbation Theory (DFPT) \cite{Campo2010, Timrov2022}, as outlined in the Methods section. 

Omitting the ${\bf k}$-point index, the general EH Hamiltonian representing the periodic material at any point of the \textbf{k}-grid can now be written as
\begin{align}
    \hat{H} = \sum_{pq,\sigma} t_{pq} \, \hat{a}_{p,\sigma}^\dagger \hat{a}_{q,\sigma}+  \sum_{p} U_p \, \hat{n}_{p,\uparrow}\, \hat{n}_{p,\downarrow} +  \sum_{pq,\sigma\tau} V_{pq} \, \hat{n}_{p,\sigma}\,\hat{n}_{q,\tau},
    \label{eq:hubbard-hamiltonian}
\end{align}
where $t_{pq}$ are hopping parameters between lattice sites, or, in other words, the localized atomic orbitals that are obtained via projection of the DFT electronic structure. Furthermore, $U_{p}$ and $V_{pq}$ are the self-consistent intra-site and inter-site Hubbard interaction parameters, respectively, $\{\hat{a}_{p\sigma},\hat{a}_{p\sigma}^{\dagger}\}$ are the fermionic operators representing an atomic orbital $p$ with spin $\sigma$, and $\hat{n}_{i\sigma} \equiv \hat{a}_{i\sigma}^{\dagger}\hat{a}_{i\sigma}$ are the respective number operators. The expression provided in \autoref{eq:hubbard-hamiltonian} serves as a general description in reciprocal space. However, in the following we will limit our investigation to the $\Gamma$-point.

For any given periodic material represented by \autoref{eq:hubbard-hamiltonian}, we are now ready to compute the direct band gap at the $\Gamma$-point by means of the ground state energies evaluated in the symmetry sectors of the Hamiltonian with $N_e-1$, $N_e$ and $N_e+1$ electrons \cite{Trushin2016}, where $N_e$ is the total number of valence electrons in the DFT pseudo-potential. More details are provided in the Methods section.

\begin{figure}[!htb] \centering\includegraphics[width=\textwidth]{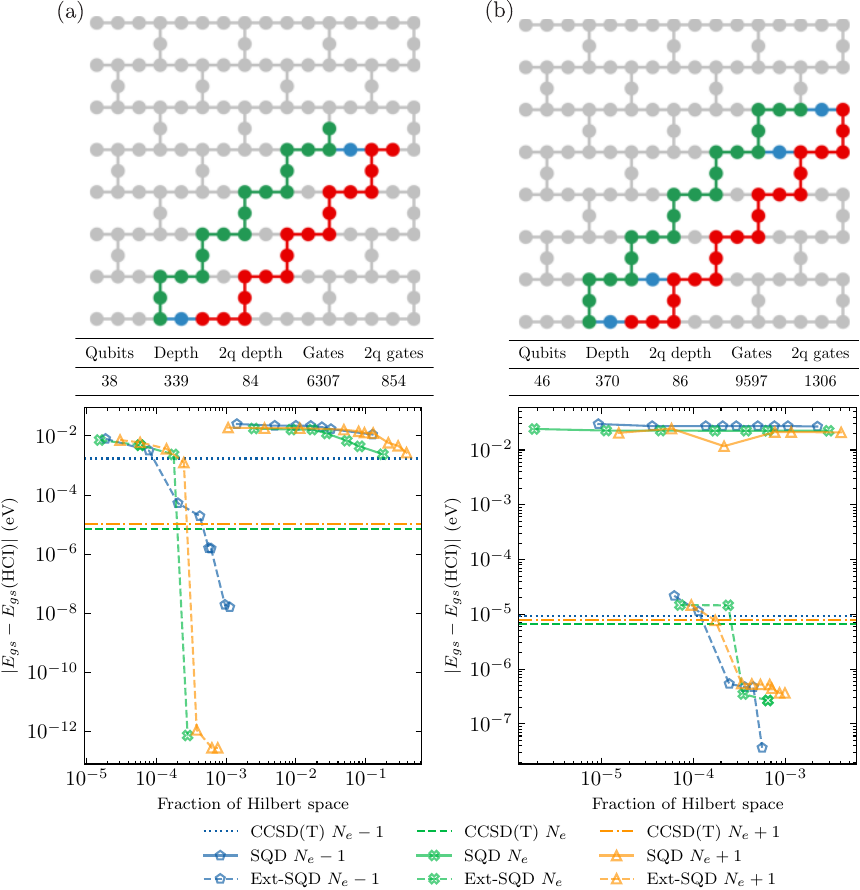} 
    \caption{SQD and Ext-SQD results obtained on the \texttt{ibm\_pittsburgh} quantum computer with circuits containing (a) 38 qubits for HfO\textsubscript{2} and (b) 46 qubits for ZrO\textsubscript{2}, respectively. The qubits representing alpha electrons are highlighted in red, the qubits representing beta electrons in green, and the qubits representing ancilla qubits in blue. (Lower panels) Error as function of configuration space fraction relative to classical diagonalization, for the electron-number subspaces $\{N_e-1, N_e, N_e+1\}$ which are needed to compute the band gap. The respective HCI energies are used as reference.
    }
    \label{fig:sqd_energies}
\end{figure}

Having established the general lattice representation for periodic materials, we will now outline the quantum computation of ground state and band gap energies of periodic materials. We implement the SQD method \cite{chemistry-paper}, as well as its extension (Ext-SQD) \cite{Barison2025}, on \texttt{ibm\_pittsburgh}, a state-of-the-art quantum system equipped with a superconducting Heron r3 quantum processing unit equipped with 156 qubits. Details with regard to algorithms and hardware specifications are provided in the Methods section. While SQD and Ext-SQD were originally designed for molecular structures, the lattice EH Hamiltonian representation in \autoref{eq:hubbard-hamiltonian} enables their application to periodic materials. As a quantum circuit approximating the true ground state's wavefunction of periodic materials, we have used a customized LUCJ ansatz \cite{lucj-ansatz} given by
\begin{equation}
    |\Psi\rangle=\prod_{\mu=1}^Le^{\hat{K}_\mu}e^{i\hat{J}_\mu}e^{-\hat{K}_\mu}|{\bf x}_{o}\rangle, \label{eq:lucj-ansatz}
\end{equation}
where $\hat{K}_{\mu} = \sum_{pq,\sigma} K_{pq}^\mu \hat{a}_{p\sigma}^{\dagger} \hat{a}_{q\sigma}$ are one-body operators that generate orbital rotations, $\hat{J}_{\mu} = \sum_{pq,\sigma \tau} J_{pq,\sigma \tau}^\mu \hat{n}_{p\sigma}^{\dagger} \hat{n}_{q\tau}$ are density-density operators between orbitals associated with adjacent qubits in Jordan-Wigner mapping, and $|{\bf x}_{o}\rangle$ is the initial Hartree-Fock (HF) state. The ansatz parameters $K_{pq}^\mu$ and $J_{pq,\sigma \tau}^\mu$ are derived from the one- and two-body amplitudes provided by CCSD.

In \autoref{fig:sqd_energies}, we plot the quantum processor layout and circuit parameters, as well as the ground state energies of HfO\textsubscript{2} and ZrO\textsubscript{2}, respectively, that were obtained obtained with the SQD and Ext-SQD methods considering both intra-site and inter-site interactions. We reference the error in the ground state energy for the three relevant symmetry sectors as a function of fraction of the configuration space used in SQD. While SQD energies are relatively far apart, the Ext-SQD values agree with the HCI reference, within a small fraction of the configuration space. For comparison, we also show the results obtained with quantum-chemical, coupled-cluster single and doubles and perturbative triples (CCSD(T)) methods which are outperformed by Ext-SQD. Considering the high accuracy obtained, we expect that the band gap predictions based on Ext-SQD will closely match the ones performed with HCI. Indeed, the difference in the band gap energies obtained with Ext-SQD and HCI is $1.6\times10^{-8} $ eV and $1.3\times10^{-7}$ eV for HfO\textsubscript{2} and Zr\textsubscript{2}, respectively. 


\begin{figure}[!htb] 
    \centering
\includegraphics[width=\textwidth]{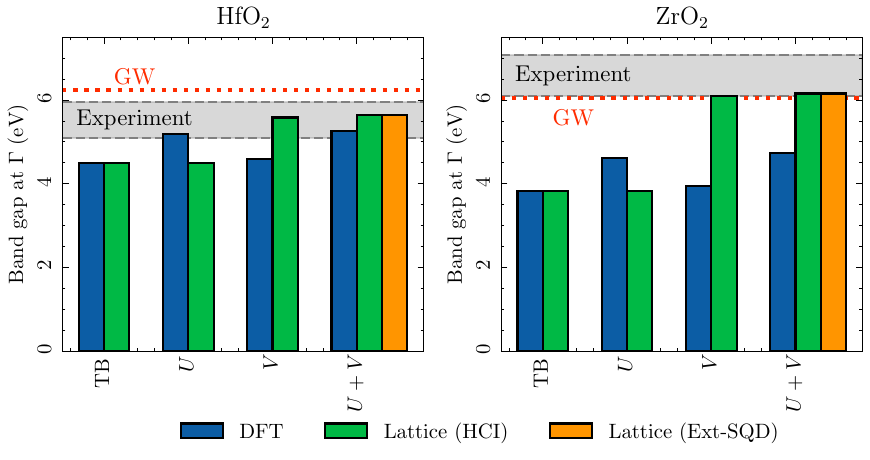} 
    \caption{Band gap energies at the $\Gamma$-point for (left) HfO\textsubscript{2} and (right) ZrO\textsubscript{2} obtained with DFT simulations, as well as with the lattice Hamiltonian representation, assuming various levels of electronic interactions: non-interacting electrons (TB), intra-site interactions ($U$), inter-site interactions ($V$), and both intra-site and inter-site interactions simultaneously ($U+V$). The Ext-SQD results were obtained with the \texttt{ibm\_pittsburgh} quantum computer. Computational results obtained in GW approximation are plotted as references. The gray shades cover the ranges of experimental band gap values as measured in the lab.}
    \label{fig:bandgap_comparison}
\end{figure}

In \autoref{fig:bandgap_comparison}, we show the quantum-computed band gap predictions, which are in agreement with the HCI benchmark, and compare them with experimental results reported in the literature.  For HfO\textsubscript{2}, we take the experimental values reported in reference \cite{jiang2010}. For ZrO\textsubscript{2}, we take the values provided in reference \cite{french1994}. For comparison, we have included both the minium and maximum band gap energies in each plot.  In both cases, the range of experimental values covered is of the order of $\sim 1\,\text{eV}$. As pointed out in reference \cite{Toshihide2005}, the variability in band gap measurements is mainly due to variations of crystal quality. For the lack of experimental errors, we consider the broad range of experimental values represented by the gray area in \autoref{fig:bandgap_comparison} as a measure of experimental variability, accounting for material imperfections and systematic measurement errors. We note that specifics with regards to the crystal phases or \textbf{k}-points probed experimentally are not consistently provided in the cited work. Experimental research results with specified crystal phase and \textbf{k}-point information could enhance the computational method development and validation in high-accuracy band gap predictions.

As a key result, we find that the quantum-computed band gap predictions based on the lattice Hamiltonian representation in \autoref{eq:hubbard-hamiltonian} for both HfO\textsubscript{2} and ZrO\textsubscript{2} outperform DFT overall. In the case of ZrO\textsubscript{2}, only the quantum-computed band gaps achieve the range of experimental values while all levels of DFT simulation fail. At the DFT level, we note that the only the intra-site interaction term $U$ leads to noticeable band gap shifts as compared to the non-interacting case. However, in the lattice representation the inter-site interaction $V$ plays the dominant role. As an additional reference, we plot band gap prediction results reported in \cite{jiang2010} which were obtained with GW approximation at the $\Gamma$-point. For both materials, the lattice representation shows prediction improvements with respect to GW. Nevertheless, GW clearly outperforms DFT+U+V in both cases. While our approach is based on a DFT calculations, it produces more accurate predictions than all levels of DFT deployed here. This could indicate that electronic correlations are indeed effectively sampled in the quantum computer while neglected in the mean-field approach. In the Supporting Information, we have provided a table with all experimental and computational reference values included for comparison.

In summary, have developed a hybrid, classical-quantum computational method for predicting band gaps of periodic materials. Using the method, we sample ground state energies of HfO\textsubscript{2} and ZrO\textsubscript{2} with minimal error mitigation overhead on a pre-fault-tolerant quantum processor. Based on the ground-state energies, we predict band gaps of the materials and compare them with predictions with standard, quantum-chemical methods as well as mean-field approaches based on density functional theory. 


The quantum-computing based predictions outperform DFT-based predictions and overlap with the range of lab experimental values reported in the literature for HfO\textsubscript{2} as well as ZrO\textsubscript{2}.  While reference simulations based on the GW approximation are close to experimental results,  they are computationally expensive \cite{Butler2016}. The quantum sampling procedure turned out to be relatively fast for the materials investigate here, producing similar results as GW with less computational resources.

In conclusion, future research should extend the application of the method to broader classes of periodic materials, as well as towards other materials properties. In terms of scalability, the main computational bottleneck is currently the classical calculation of configuration interaction. With current advances in the development of quantum processors, we expect that the quality quantum sampling should significantly improve even with noisy intermediate-scale quantum processors.

\section*{Methods}

\subsection*{Density functional theory and lattice projection} \label{sec:dft-methods}

For obtaining the electronic density of the ground state based on density functional theory (DFT) \cite{Hohernberg1964, Kohn1965}, we used the Quantum Espresso (QE) software package \cite{Gianozzi2009,Gianozzi2017,quantum_espresso}. The materials' geometries were optimized assuming a cubic lattice for both ZrO\textsubscript{2} and HfO\textsubscript{2}. The electronic valence configurations were $4s^2 4p^6 5p^0 4d^2 5s^2$ for Zr and $5s^2 5p^6 6s^2 5d^2$ for Hf and $2s^2 2p^4$ for O, respectively. The pseudopotential for O was taken from the SSSP PBE Efficiency v1.3.0 library \cite{Prandini2018}. For the transition metal the choice depended on the tight-binding basis length and we selected  \texttt{Zr.pbe-spn-rrkjus\_psl.1.0.0.UPF} and \texttt{Hf-sp.oncvpsp.UPF}, respectively. The kinetic energy cutoffs for the wavefunctions, the charge densities and the \textbf{k}-grids were converged for each material, leading to material-specific parameter sets.


We computed the intra-site $U$ and inter-site $V$ Hubbard parameters with QE in the self-consistent DFT+U+V framework following references \cite{Campo2010,Timrov2022}. Initial guesses for $U$ and $V$ of $10^{-8}~\text{eV}$ were provided as input for the self-consistent step, using orthogonalized atomic orbitals as Hubbard projectors. We considered $U_{4d}$ and $V_{4d\text{-}2p}$ for ZrO\textsubscript{2} and $U_{5d}$ and $V_{5d\text{-}2p}$ for HfO\textsubscript{2}, respectively. The optimized lattice parameters, along with the $U$ and $V$ values for each material,  are provided in the Supporting Information. The lattice constants are in agreement with experimental results \cite{Bergamin1997, Muller1968, Wang1992} and the $U$- and $V$-values are consistent with references \cite{Gebauer2023, Hall2021, Yang2025}.

The tight-binding representation was calculated by a projection of the plane-wave electronic structure from QE onto localized atomic orbitals using the PAOFLOW software \cite{paoflow, Cerasoli2021}. The projection outcome is a representation of the electronic structure as a lattice Hamiltonian, expanded in the atomic orbital basis present in the pseudopotential, with its respective coefficients, i.e., hopping parameters. The QE output reflects standard DFT, i.e., without  $U$ and $V$ corrections. We found that the tight-binding projection described the materials reasonably well, reproducing the DFT energies of their respective band structures, as shown in the Supporting Information. Note, that the hopping parameters and the inter-site interactions are not restricted to first or second neighbors in the lattice. Rather, they are determined by the DFT projection and, therefore, specific to each material. A non-zero hopping term between two orbitals means that they are considered ``neighbors''.

\subsection*{Mapping between lattice and electronic Hamiltonian}\label{sec:diagonalizing_lattice_hamiltonian}

The ground state energy of the extended Hubbard Hamiltonian described in \autoref{eq:hubbard-hamiltonian} was computed using the SQD framework originally designed for quantum chemistry calculations of isolated molelcules \cite{chemistry-paper}. In that context, the system is generally represented by using a general electronic Hamiltonian in second quantization of the form
\begin{align}
    H_{chem} = \sum_{pq,\sigma} h_{pq}  \, c_{p,\sigma}^\dagger c_{q,\sigma} +  \frac{1}{2}\sum_{pqrs,\sigma\tau} h_{pqrs,\sigma\tau} \, c_{p,\sigma}^\dagger c_{r,\tau}^\dagger c_{s,\tau} c_{q,\sigma}, \label{eq:molecular-hamiltonian}
\end{align}
where $h_{pq}$ and $h_{pqrs,\sigma \tau}$ are the one- and two-body tensors derived from the overlap and electronic repulsion integrals, respectively. Within the two-body tensor, for some fixed spatial orbital indices $pqrs$ the tensor element $h_{pqrs}$ can have different values depending on spins orientation which is captured in $h_{pqrs,\sigma\tau}$. 

For the purpose of this investigation, we expressed the extended Hubbard model in \autoref{eq:hubbard-hamiltonian} as a molecular Hamiltonian by means of the following mappings:
\begin{align}
    h_{pq} &\leftrightarrow t_{pq}, \nonumber\\
    h_{pppp,\sigma \tau} &= 2U_p \nonumber\\
    h_{ppqq,\sigma \tau} &= 2V_{pq} \\
    h_{pppp,\sigma \sigma} &= 0 \nonumber\\
    h_{pqrs,\sigma \tau} &= 0 \:\:\: \text{if}\:\:\: p\neq q \:\:\:\text{or}\:\:\: r\neq s. \nonumber \label{eq:hubbard_to_molecular}
\end{align}

\subsection*{Sample-based quantum diagonalization} \label{sec:sqd-methods}

For diagonalizing the lattice Hamiltonian in \autoref{eq:hubbard-hamiltonian}, we used the sample-based quantum diagonalization method, SQD, as well as its extended version, Ext-SQD. In a first step, we mapped the respective hopping and Hubbard interaction parameters to one- and two-body integrals of an effective electronic Hamiltonian, see Supporting Information, enabling the application of quantum-chemical methods for diagonalization, i.e., CCSD, CCSD(T), as well as HCI. In a second step, we performed a CCSD calculation using PySCF package \cite{pyscf}. Using the one- and two-body amplitudes as input, we constructed a single layer, LUCJ ansatz ($\mu=1$ in \autoref{eq:lucj-ansatz} by using \texttt{ffsim} \cite{ffsim}.

We executed the parametrized circuits on a quantum processor to collect a set of samples given by $\{ |{\bf x}\rangle \}$, where ${\bf x}$ are bitstrings. For each of the calculations, we performed a total of $2.5 \times 10^6$ shots on the device. Following the procedure outlined in reference \cite{chemistry-paper}, we constructed a set $\mathcal{S}^{(d)}$ of $d$ electronic configurations, where $d$ is determined implicitly by means of the fraction of the configuration space shown in \autoref{fig:sqd_energies} . We then diagonalized the Hamiltonian projected onto the $d$-dimensional subspace spanned by the configurations in the set. The projected Hamiltonian can be written as
    \begin{equation*}
        \hat{H}_{\mathcal{S}^{(d)}} = \hat{P}_{\mathcal{S}^{(d)}} \hat{H} \hat{P}_{\mathcal{S}^{(d)}},\:\: \text{ with } \hat{P}_{\mathcal{S}^{(d)}} = \sum_{{\bf x}\in \mathcal{S}^{(d)}} |{\bf x}\rangle\langle{\bf x}|.
    \end{equation*}
By increasing $d$ stepwise, we performed the classical diagonalization in larger configuration spaces, leading to a trade-off between accuracy  and computational cost.
    
The Ext-SQD algorithm \cite{Barison2025} extends the configuration space obtained from an initial SQD calculation. In a first step, we removed the ground state configurations whose probabilities are below a certain threshold. We then applied excitation operators to the reduced set for obtaining and extended configuration set. In a next step, we classically compute the ground state on this extended configuration space for obtaining a  refined ground-state energy. As shown in \autoref{fig:sqd_energies}, Ext-SQD improved the energy accuracy relative to SQD in smaller configuration spaces, thus reducing computational costs. Throughout this work, we used the SQD result with the largest configuration space as input for Ext-SQD calculations. In the case of HfO\textsubscript{2}, we performed Ext-SQD by applying single excitations to the original set of configurations, with a threshold of $1.0\times10^{-4}$. For ZrO\textsubscript{2}, we have included single and double excitations and the threshold was set to $2\times10^{-5}$.
    
For the results shown in \autoref{fig:sqd_energies}, we considered various values of $d$, or, in other words, various sizes of the configuration space. We used the SQD and HCI workflows available in \texttt{qiskit-addon-sqd} and \texttt{qiskit-addon-dice-solver} software packages. The diagonalizations with regard to SQD, Ext-SQD and HCI were performed using the Davidson algorithm within PySCF \cite{pyscf} and DICE \cite{DICE}. In our quantum-processor based simulations, we used Dynamical Decoupling as error mitigation method \cite{dynamical-decoupling}, with standard configurations as implemented in the Qiskit library version 1.3.0~\cite{qiskit}.

\subsection*{Band gap calculation} \label{sec:bandgap-methods}

Having determined a material's ground state energies with SQD and Ext-SQD, respectively, we estimated its band gap as a function of the ground-state energy of the Hamiltonian in the $N_e$, $N_e-1$ and $N_e+1$ electron number symmetry sectors based on $E_g = E[N_e-1] + E[N_e+1] - 2E[N_e]$ \cite{Trushin2016}, where $E[N]$ is the ground-state energy of the Hamiltonian in the $N$ electron symmetry sector. For comparison, the DFT band gap was calculated as the energy difference between the lowest unoccupied and highest occupied Kohn-Sham states in the band structure. We validated the DFT-derived band gaps with the results obtained using the tight-binding lattice Hamiltonian, without inclusion of Hubbard parameters. 

\bmhead{Code Availability}
The code used to generate the results reported in the current study is available as a Github repository, at \href{https://github.com/neumannrf/sqd-band-gaps}{https://github.com/neumannrf/sqd-band-gaps}.
Open-source software used in this work includes Quantum Espresso \cite{quantum_espresso}, PAOFLOW \cite{paoflow}, PySCF \cite{pyscf}, \texttt{ffsim} \cite{ffsim}, Qiskit SDK \cite{qiskit}, \texttt{qiskit-addon-sqd} \cite{qiskit-addon-sqd}, \texttt{qiskit-addon-dice-solver} \cite{qiskit-addon-dice-solver}, \texttt{qiskit-addon-aqc-tensor} \cite{qiskit-addon-aqc-tensor}.

\bmhead{Data Availability}
The datasets used and analyzed during the current study are available from the corresponding author on reasonable request.

\bmhead{Acknowledgments}

We thank Mario Motta, Mirko Amico, Kevin Sung, Petar Jurcevic, Ieva Liepuoniute, Tanvi Gujarati, Aleksandros Sobczyk, Ed Chen, Gavin Jones and Javier Robledo Moreno (IBM) for discussions and expert technical and project support. We acknowledge Ramon Cardias (CBPF) for introducing the PAOFLOW software to us and Prof. Marcio Costa (UFF) for guidance on the use of PAOFLOW. One author received funding by the NCCR Catalysis (grant numbers 180544 and 225147), a National Centre of Competence in Research funded by the Swiss National Science Foundation.

\bibliography{sn-bibliography}

\newpage
\section*{Supporting Information}

\renewcommand{\theequation}{S\arabic{equation}}
\renewcommand{\thefootnote}{\alph{footnote}}
\renewcommand{\thesection}{S\arabic{section}}
\renewcommand{\thefigure}{S\arabic{figure}}
\renewcommand{\thetable}{S\arabic{table}}
\setcounter{equation}{0}
\setcounter{section}{0}
\setcounter{figure}{0}
\setcounter{table}{0}
\setcounter{page}{1}

\subsection*{Electronic structure and tight-binding projection}

We also applied our lattice Hamiltonian description to the Si bulk crystal. Campo \textit{et al.} argued that Silicon bulk is a ``strongly hybridized'' system, where \textit{s} and \textit{p} orbital hybridization, inter-site and intra-site, plays an important role for the electronic structure \cite{Campo2010}. The authors also have shown a good agreement to experimental results regarding the band gap prediction through the addition of $U$ and $V$ values calculated self-consistently. 
An optimized cubic structure was considered, with an electronic valence configuration of $3s^2 3p^2 3d^0$. We used a Perdew-Burke-Ernzerhof (PBE) norm-conserving Martins-Troullier GIPAW \cite{Tantardini2022} pseudopotential available in the QE pseudopotentail database \cite{qe_pseudopotentials} as \texttt{Si.pbe-mt\_gipaw.UPF}, with a kinetic energy cutoff for wavefunctions and charge density of 50 Ry and 200 Ry, respectively. It is worth noting that the inclusion of \textit{d} orbitals in the valence bands was essential to later perform the tight-binding projection. 


The band gap results for Si are shown in \autoref{tab:bandgap-results}. We can see that according to our results, from both DFT and HCI values, the addition of electronic interaction does not improve the prediction towards the experimental value, and the difference between the predictions from DFT and the lattice representation is mild. It is important to highlight here, that while the inter-site $V$ interaction between \textit{s} and \textit{p} orbitals was added, intra-site $V$ couplings were not included since the methodology used here does not allow the computation of $V$ values for intra-site manifolds. The lack of this coupling can indicate why the band gap was not improved in our work, differently from the work of Campo \textit{et. al.} \cite{Campo2010}.

The computed direct band gap values at $\Gamma$ for all materials considered in this work are shown in \autoref{tab:bandgap-results}, along with previous theoretical and experimental results from the literature. We highlight that the results shown here are the also shown in \autoref{fig:bandgap_comparison}.


\begin{table}[ht]
\centering
\caption{Direct band gaps of ZrO\textsubscript{2}, HfO\textsubscript{2}, and Si bulk (in units of eV). The values provided for the tight-binding (TB) and Extended Hubbard model (EHM) were computed using the HCI method.}
\begin{tabular*}{0.7\textwidth}{@{\extracolsep{\fill}}lccc}
\toprule
Method & ZrO\textsubscript{2} & HfO\textsubscript{2} & Si bulk\\
\midrule
\multicolumn{4}{l}{\textit{This work}} \\
DFT (GGA)       & 3.83 & 4.50 & 2.57 \\
DFT+U+V         & 4.74 & 5.27 & 2.53 \\
TB   & 3.83 & 4.50 & 2.57 \\
EHM  & 6.16 & 5.64 & 2.53 \\
\midrule
\multicolumn{4}{l}{\textit{Other works}} \\
DFT             & 3.65$^{a}$ (LDA) & $\sim$ 4.2$^{b}$ (LDA) & 2.57$^{c}$ (LDA) \\
GW              & 5.81$^{a}$ & $\sim$ 6$^{b}$ (GW$_0$) & 3.27$^{c}$ \\
Experimental    & 6.1-7.08$^{e}$ & $5.10-5.95^b$ & 3.40$^{c}$ \\
\bottomrule
\label{tab:bandgap-results}
\end{tabular*}

\vspace{0.5em}
\footnotesize
$^{a}$ Ref. \cite{kralik1998}
$^{b}$ Ref. \cite{jiang2010}
$^{c}$ Ref. \cite{Hybertsen1985}
$^{e}$ Ref. \cite{french1994}
\end{table}

The computation of $U$ and $V$ values should be converged with respect to the Monkhorst-Pack grid \textbf{q}-mesh, similarly to the \textbf{k}-mesh in the ground state calculation. Here, we considered that $q = k/2$ as sufficient. Larger \textbf{q}-mesh grids were tested for convergence, resulting in a variation of the order of 0.01 and 0.001 eV for $U$ and $V$ values, respectively. Similar results have been found in previous works \cite{Hall2021}.
Only nonequivalent atoms by symmetry were perturbed 
and the convergence threshold for the response function $\chi$ was set as $10^{-7}$.
In \autoref{tab:dft-values}, the optimized lattice constants and self-consistent $U$ and $V$ Hubbard parameters are shown for the materials considered.


\begin{table}[ht] 
\centering
\caption{Optimized lattice constants and self-consistent $U$ and $V$ Hubbard parameters for selected materials.}
\begin{tabular}{lccc}
\hline\hline
Material & Lattice constants (Å) & U (eV) & V (eV) \\
\hline
Si bulk   & a = 5.45 & 2.42 (Si$_{3p}$) and 1.91 (Si$_{3s}$)  & -0.09 (Si$_{3p}$–Si$_{3s}$) \\
ZrO\textsubscript{2}   & a = 5.12 & 3.13 (Zr$_{4d}$) & 0.64 (Zr$_{4d}$–O$_{2p}$) \\
HfO\textsubscript{2}   & a = 5.08 & 2.66 (Hf$_{5d}$) & 0.57 (Hf$_{5d}$–O$_{2p}$) \\
\hline\hline
\label{tab:dft-values}
\end{tabular}
\end{table}

\begin{figure}[!htb]
    \centering
    \includegraphics[width=1.\textwidth]{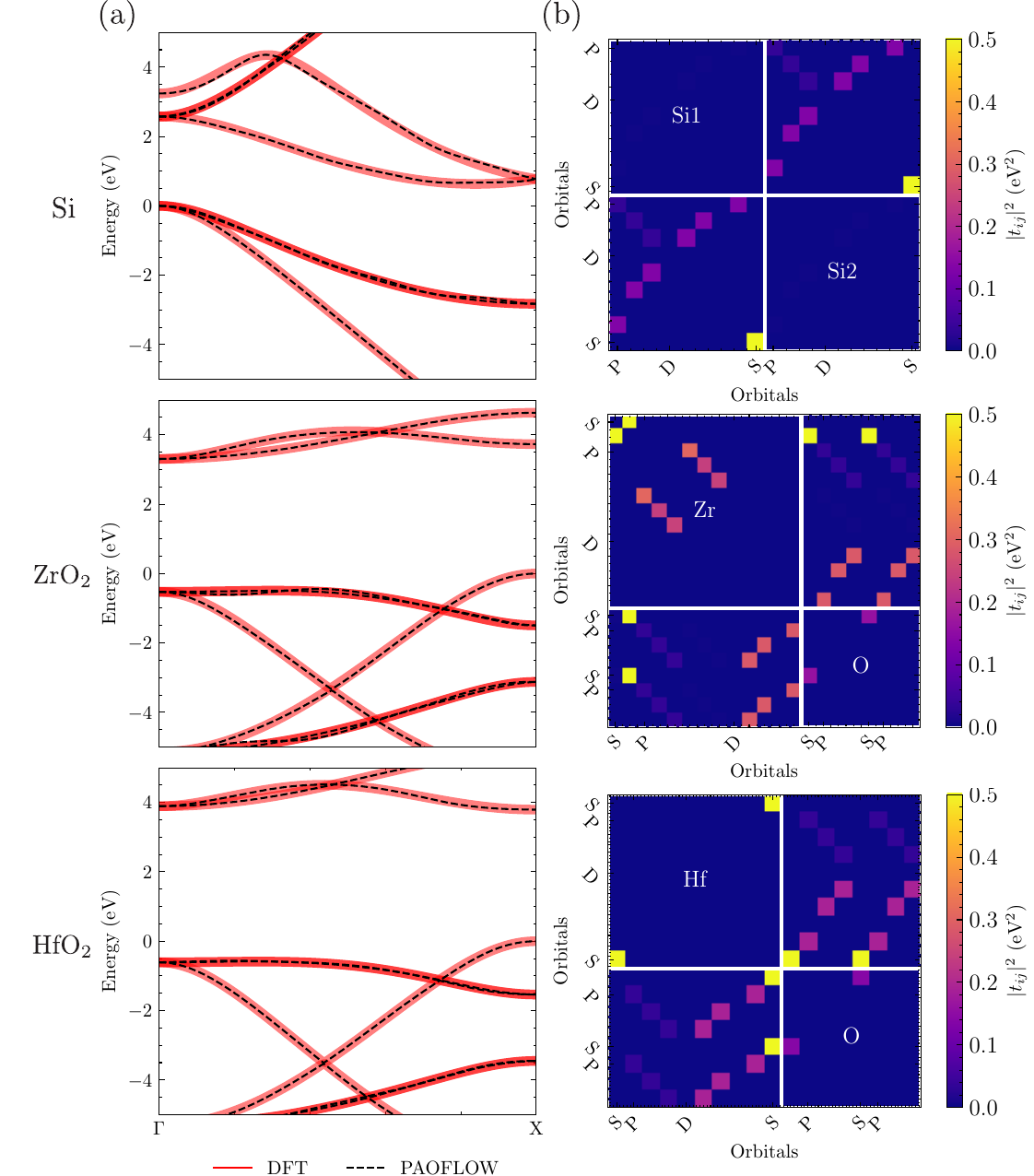} 
    \caption{Results obtained from the projection of the DFT electronic structure. (a) Band structures for Si (top), ZrO\textsubscript{2} (middle) and HfO\textsubscript{2} (bottom). The Quantum Espresso results using standard DFT are shown in dashed black line, while the PAOFLOW projection is represented by solid red lined. (b) Graphical representation of the Hamiltonian at the $\Gamma$-point for the same materials. The color map represents the normalized squared value of the hopping matrix elements ($|t_{ij}|^2$) and the columns and rows are the localized orbital atomic basis. To allow for a good visualization of the interaction and hopping terms, we do not show the diagonal elements, which are associated with the energy of orbitals.}
    \label{fig:SI_bandstructure}
\end{figure}

A comparison between the Quantum Espresso \cite{Gianozzi2009,Gianozzi2017,quantum_espresso} and PAOFLOW \cite{paoflow, Cerasoli2021} band structures of each semiconductor investigated in this work is presented in \autoref{fig:SI_bandstructure}(a). This comparison indicates an excellent performance of tight-binding projection executed by PAOFLOW, where the occupied and first unoccupied bands are accurately represented. Note that the match between DFT and PAOFLOW bands could be further enhanced, however, this is outside the scope of this work, which considers only the $\Gamma$-point. 

A graphical visualization of the tight-binding Hamiltonian can be seen in \autoref{fig:SI_bandstructure}(b). The color map represents the hopping parameters, indicating the interaction among atomic orbital sites. The matrix is divided in blocks, according to the site of each atom of the unit cell. Therefore, interactions within the diagonal and off-diagonal blocks are intra-site energies and inter-site hopping parameters, respectively. 

\subsection*{HCI as a classical benchmark} \label{sec:hci_benchmark}

In the main text we consider HCI \cite{HCI} as the ground truth for comparing the ground state energies obtained with SQD and Ext-SQD, even though it is an approximate method. In this section we show why that is the case, at least for the systems analyzed in this work. 

\begin{figure}[!htb] 
    \centering
\includegraphics[width=\textwidth]{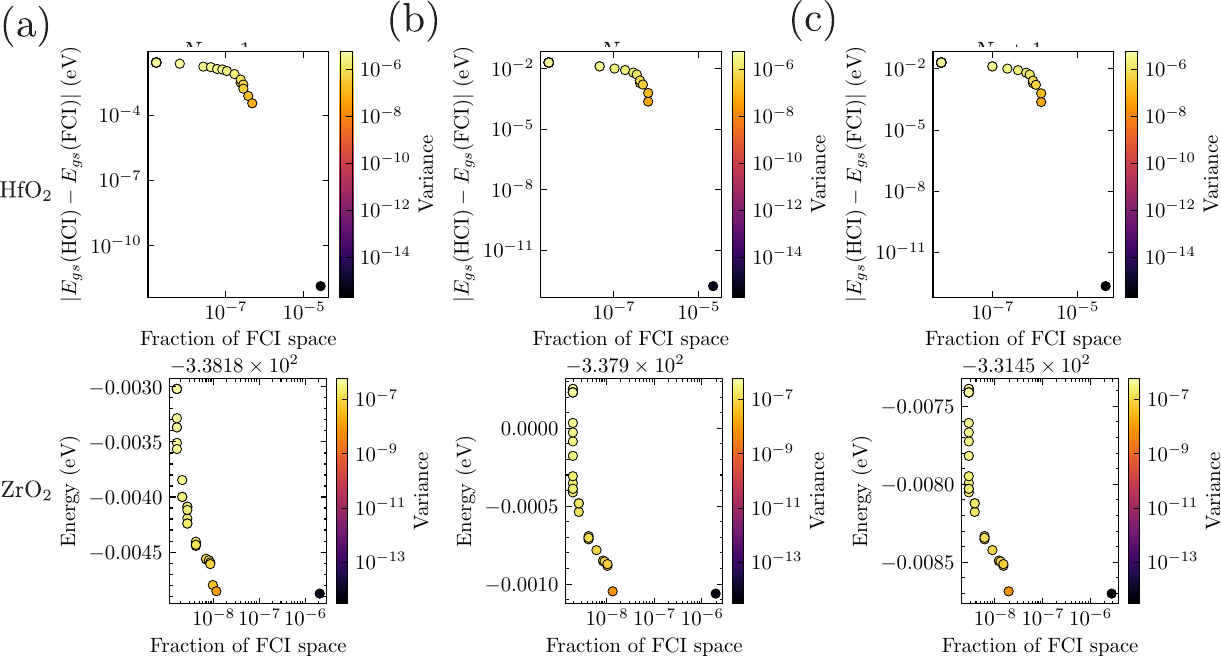} 
    \caption{HCI results for the ground state energy of HfO\textsubscript{2} and ZrO\textsubscript{2} for the $N_e$ (a), $N_e-1$ (b) and $N_e+1$ (c) electron number subspaces. For HfO\textsubscript{2} we show the error in respect with FCI (exact), and for ZrO\textsubscript{2} we show the actual value of the energy. In the horizontal axis we show the fraction of the total configuration space considered in the calculation of the ground state, in a similar fashion as in \autoref{fig:sqd_energies}. We also add color bars to show the variance of the ground state obtained in each calculation, and we define the variance as $\left(\langle H^2\rangle -\langle H \rangle^2\right)/\langle H \rangle^2$.}
    \label{fig:hci_plot.pdf}
\end{figure}

In \autoref{fig:hci_plot.pdf}, we show the HCI results for the ground state energies of selected materials, in the three relevant electron-number subspaces. For HfO\textsubscript{2} (top) we show the error in the energy with respect to the exact solution given by Full Configuration Interaction (FCI) as implemented in PySCF \cite{pyscf}. For ZrO\textsubscript{2} (bottom), which has a configuration space too large to allow for an exact solution, we show the actual value of the energy instead. We also indicate the variance of the ground state obtained with HCI, and we highlight that a small variance is an indicator of high quality of the approximation of the ground state, since any eigenstate of a Hamiltonian should have zero variance. We can see that for all cases the variance obtained with HCI (for the largest considered configuration space) is on the order of $10^{-15}$ (machine precision).

We can see that for HfO\textsubscript{2} the ground state energy given by HCI matches the exact one up to machine precision, which justifies the usage of HCI as a benchmark for the ground truth. For ZrO\textsubscript{2}, we cannot make the same comparison because we do not have the exact energy, but the small value of the variance obtained indicates that the ground state obtained with this method closely matches the exact ground state. With these results we can be certain about the success of HCI in obtaining the ground state energy for the systems considered.

\subsection*{Reference Table} \label{sec:reftable}

In \autoref{tab:bandgap}, we show a summary of band gap predictions taken from the literature (at $\Gamma$) for ZrO\textsubscript{2}, alongside values obtained from our method (i.e., the HCI values in \autoref{fig:bandgap_comparison}). 

\begin{table}[ht]
    \centering
    \begin{tabular}{l l c} 
        \toprule
        Method & & Direct band gap (eV) \\
        \midrule
        OLCAO \cite{french1994} & Theory & 4.9 \\
        LDA \cite{kralik1998} & Theory & 3.6 \\
        GW \cite{kralik1998} & Theory & 5.8 \\
        OLCAO \cite{zandiehnadem1988} & Theory &  3.8 \\
        DFT \cite{dash2004} & Theory & 3.9 \\
        \bf{This work} & Theory & \bf{6.1} \\
        Vaccum UV Spectroscopy\cite{french1994} & Experiment & 6.1,7.1 \\
        \bottomrule
    \end{tabular}
    \caption{Direct band gap of ZrO\textsubscript{2} at the $\Gamma$  point as obtained by previous works, compared this work. For our contribution we consider the value obtained from HCI, also shown in \autoref{fig:bandgap_comparison}.}
    \label{tab:bandgap}
\end{table}

\end{document}